\documentclass{llncs}

\usepackage{paralist,fancybox,listings, graphicx, algorithm2e,caption}

\begin{document}

\title{From Pretty Good To Great: Enhancing PGP using Bitcoin and the Blockchain (FULL VERSION)}
\titlerunning{Hamiltonian Mechanics}  
%
\author{Duane Wilson\thanks{Department of Computer Science, Johns Hopkins University} and Giuseppe Ateniese\thanks{Department of Computer Science, Sapienza University of Rome}}
\authorrunning{Ivar Ekeland et al.} 
%
\institute{}

\maketitle              

\begin{abstract}
PGP is built upon a Distributed Web of Trust in which a user's trustworthiness is established by others who can vouch through a digital signature for that user's identity.  Preventing its wholesale adoption are a number of inherent weaknesses to include (but not limited to) the following: \textbf{1)} Trust Relationships are built on a subjective honor system, \textbf{2)} Only first degree relationships can be fully trusted, \textbf{3)}  Levels of trust are difficult to quantify with actual values, and \textbf{4)} Issues with the Web of Trust itself (Certification and Endorsement).  Although the security that PGP provides is proven to be reliable, it has largely failed to garner large scale adoption.  In this paper, we propose several novel contributions to address the aforementioned issues with PGP and associated Web of Trust.  To address the subjectivity of the Web of Trust, we provide a new certificate format based on Bitcoin which allows a user to verify a PGP certificate using Bitcoin identity-verification transactions - forming first degree trust relationships that are tied to actual values (i.e., number of Bitcoins transferred during transaction).  Secondly, we present the design of a novel Distributed PGP key server that leverages the Bitcoin transaction blockchain to store and retrieve Bitcoin-Based PGP certificates.  Lastly, we provide a web prototype application that demonstrates several of these capabilities in an actual environment.
\end{abstract}
In a recent article, Yahoo announced its intentions to add an extension that will provide its customers with the ability to digitally sign and encrypt messages using Pretty Good Privacy (PGP).  Yahoo plans to use a fork of Google's End to End OpenPGP plugin that is currently in development.  Yahoo follows the likes of Google, Facebook and Microsoft, who also recently announced they would encrypt internal traffic in response to the Snowden spying revelations \cite{yahoo}.  Traditional methods of securely sharing between two or more parties rely on the use of Public-Key Encryption within a Public Key Infrastructure (PKI).  In a traditional PKI scheme, a certificate authority or certification authority (CA) is an entity that issues digital certificates. The digital certificate certifies the ownership of a public key by the named subject of the certificate. This allows others (relying parties) to rely upon signatures or assertions made by the private key that corresponds to the public key that is certified. In this model of trust relationships, a CA is a Trusted Third Party (TTP) that is trusted by both the subject (owner) of the certificate and the party relying upon the certificate. CAs are characteristic of many PKI schemes. \cite{ca}.   Currently, the most viable alternative for Public Key Crytography based on a CA is PGP.  PGP is built upon a Distributed Web of Trust in which a user's trustworthiness is established by others who can vouch for that user's identity.  Preventing its wholesale adoption are a number of inherent weaknesses to include (but not limited to) the following: 

\begin {itemize}
\item Trust Relationships are built on a subjective honor system  
\item Only first degree relationships can be fully trusted
\item Levels of trust are difficult to quantify with actual values 
\item Issues specific to the Web of Trust:\\ 
\begin {enumerate}
\item \textbf{Certification.} It is currently difficult to get certified if the key is new. In general people complain that it is hard to find endorsers to enhance the trustworthiness of a new key - which will limit its use.
\item \textbf{Endorsement.} Competence and willingness of endorsers.  There is currently no incentive to endorse a key of someone you know - much less someone you indirectly know through a friend or relative.
\end {enumerate}
\end {itemize}

Bitcoin is a form of digital currency, created and held electronically \cite{cd}.  According to ``Crypto Coin News'', the number of active Bitcoin users worldwide will reach 4.7 million by the end of 2019, marking a significant gain over the 1.3 million last year, according to a report from Juniper Research \cite{busers}.  As a result of its popularity, we introduce a new Bitcoin-Based PGP certificate format, certificate validation methodology, and certificate endorsement model that overcomes the issues we have highlighted above.  Issues 1 and 2 with the Web of Trust can be easily solved using our new Bitcoin-Based PGP certificate format and verification system. Issue 4 can be resolved by use of endorsement fee. The amount of the fee can be determined by the user and will vary based on the current value of a Bitcoin - which has been relatively stable of late \cite{stable}.  In Issue 2, the bitcoin payment ensures that the endorser follows the ``authentication'' procedure otherwise they risk losing bitcoins - which demonstrates both their competence and willingness to serve as a viable certificate endorser.

There are some interesting properties of our trust establishment protocol that could result in safer use of PGP. Property 1) People have the option of using previous transactions before using a certificate OR directly establishing a trust relationship themselves with a certificate owner (i.e., direct trust). Property 2) As mentioned above, because of the risk of losing bitcoins via the identity-verification process, people will be less likely to leverage our certificates without a direct trust establishment.  Property 3) The block chain and associated identity-verification transactions provide transparency into the trustworthiness of others.  In addition to these benefits, we also provide the design of a novel PGP Key Server based on the blockchain's ability to store pieces of data since the 0.9.0 release.  The 0.9.0 release of Bitcoin Core added a new standard transaction type granting access to a previously disallowed script function, \textit{OP-RETURN} \cite{opreturn}.  This function accepts a user-defined sequence of up to 40 bytes (now 80 bytes in current release).  Once realized, this new key server will be completely de-centralized and serve as an appropriate repository for Bitcoin-Based PGP Certificates.  Our work specifically provides the following contributions:

\begin{itemize}
\item \textbf{Bitcoin-Based PGP Certificate: }  Contains Bitcoin address for identity verification and certificate revocation.   
\item \textbf{Identity-Verification and Revocation Transactions: } Serves as alternative means of verifying a certificate owner's Public Key contained in a Bitcoin-Based PGP Certificate.  Also provides a mechanism for certificate revocation using the embedded Bitcoin address for revocation purposes.
\item \textbf{PGP Trust Levels: } Allows users to specify the
amount of Bitcoins they are willing to ``risk'' in order to verify a particular Bitcoin-Based PGP certificate.  This amount (determined by the verifier) now attests to level of trust the verifier has in the certificate owner. 
\item \textbf{Certificate Signing Endorsements: } Adds small incentive fee (via Bitcoin) each time an endorser (with a valid Bitcoin address) signs an Enhanced PGP Certificate stored on Bitcoin-Based PGP Key Server. 
\item \textbf{Bitcoin-Based PGP Key Server Design: }  Demonstrates method of using the Bitcoin Transaction Blockchain for PGP Key Storage.  Offers a completely decentralized client-based software application that allows users to efficiently store and retrieve Bitcoin-Based PGP certificates within the blockchain.  Application will separate certificate into individual pieces to fit within the allowed number of bytes and store within blockchain - as append only data.  Similarly, upon request (e.g., based on PGP Key ID or similar), client will facilitate the retrieval of PGP certificate fragments and reassemble them for use by requesting user.   
\end{itemize}

The rest of this paper is organized as follows: Section 2 discusses the work related to this area of research, Section 3 provides a background of the current PGP Public Key Certificate format as a context for our Bitcoin-Based PGP certificate, Section 4 presents an overview of PGP threats addressed by our contributions, Section 5 discusses the design details of our prototype application and new key server, Section 6 provides relevant sample output from the primary prototype functions, and Section 7 concludes the paper and identifies areas for future work. 

\section{Related Work}
This section examines previous work that proposes novel methods to protect the confidentiality of data in an uncontrolled network-accessible environment without the aid of a Certificate Authority (CA).  According to \cite{bitpay1}, BitPay has launched a project that leverages bitcoin technology to facilitate a decentralized authentication system.  Called BitAuth, the system uses cryptographic signatures in place of server-side password storage. BitAuth is a way to do secure, passwordless authentication using the same elliptic-curve cryptography as Bitcoin. Instead of using a shared secret, the client signs each request using a private key and the server checks to make sure the signature is valid and matches the public key. A nonce is used to prevent replay attacks and provide sequence enforcement \cite{bitpay2}.  Similar to our novel Bitcoin-Based PGP certificate, BitAuth provides ``portable'' identity in that the same identity can be used with (or on) multiple services. BitAuth is a promising new method of authentication, but currenly relies heavily on the System Identification Number (SIN).  The SIN is a new concept that is similar to a Bitcoin address, however, is not widely adopted.  Whereas, our scheme relies on popular Bitcoin primitives - address, transactions, and the block chain - that are widely being used.  Additionally, since the focus of BitAuth is on authentication, it cannot be used to protect the confidentiality of information shared between two parties - as is the primary benefit of our Bitcoin-Based PGP Certificate.\\  
\indent Off-the-Record (OTR) Messaging is a protocol designed for private social communications. According to \cite{otr1,otr2}, the notion of an off-the-record conversation captures the semantics one intuitively wants from private communication - only the two parties involved are privy to the contents of the conversation; after the conversation is over, no one (not even the parties involved) can produce a transcript; and although the participants are assured of each other's identities, neither they nor anyone else can prove this information to a third party. Current versions of the OTR protocol, support mutual authentication of users using a shared secret through the socialist minimalist protocol.  This feature makes it possible for users to verify the identity of the remote party and avoid a man-in-the-middle attack without the inconvenience of manually comparing public key fingerprints through an outside channel.  OTR's primary weakness lies in the fact that it is primarily applicable in the domain of instant messaging - whereas our Bitcoin-Based PGP certificate can be used in virtually any domain in which secure information sharing is desired.  According to the authors of the OTR protocol, ``The high latency of email communication makes using our``off-the-record'' protocol impractical in the setting of email.'' \\
\indent In \cite{convergence}, a secure replacement for CAs is proposed.  Rather than employing a traditionally hard-coded list of immutable CAs, Convergence allows one to configure a dynamic set of Notaries which use network perspective to validate user communications.  It provides the following guarantees: Trust Agility, Robustness, Backwards Compatibility, Extensibility and Anonymity.  This all occurs within a distributed environment in which anyone can serve as a trust notary. Convergence originated from the ideas originally developed by the Perspectives Project at Carnegie Mellon University \cite{perspectives}.  Convergence has great promise in the domain of web browser security and other areas where SSL is prominent.  However, it suffers from the fact that the number of notaries currently in existence for performing CA functions is limited (due to it being a fairly new effort).  As a result, this could lead to potential Denial of Service (DoS) attacks in the event the notaries become overwhelmed with requests.  The Bitcoin infrastructure - upon which our certificate primarily relies - has successfully processed nearly 40 million transactions (to date) \cite{stats}.  This makes it robust against volume-based security attacks such as DoS and DDoS - when applicable.   

\section{Public-Key Digital Certificates}
In this section we discuss the differences between the traditional PGP Certificate and our novel Bitcoin-Based PGP certificate which leverages aspects of the Bitcoin infrastructure to address the shortcomings of existing PGP certificate web of trust model. 

\subsection{PGP Certificates}
PGP is a public key cryptographic package, which is intended for public usage. It provides sender authenticity, message integrity and non-repudiation of the sender. Although PGP can encrypt any data or files, it is most commonly used for e-mail which has no built-in security as originally implemented. It was  originally designed and developed by Phil Zimmermann in 1991.  A PGP certificate includes (not limited to) the following information \cite{pgpworks}:
\begin {itemize}
\item \textbf{PGP Version Number}: This identifies which version of PGP was used to create the key associated with the certificate. 
\item \textbf{Certificate holder's public key}: This is the public portion of the key pair, together with the algorithm of the key: RSA, Elgamal or DSA.   
\item \textbf{Certificate holder's information}: This is information about the user, such as his or her name, user ID, e-mail address, ICQ number, photograph, or other identifier.
\item \textbf{Digital signature of the certificate owner}: This item, also called a self-signature, is the signature generated using the corresponding private key of the public key associated with the certificate. 
\item \textbf{Validity Period}: The certificate's start date/time and expiration date/time; indicates when the certificate will expire. If the key pair contains subkeys, then this includes the expiration of each of the encryption subkeys as well. Subkeys enable convenient use of separate keys for signing and encryption. 
\item \textbf{Preferred symmetric encryption algorithm for the key}: This indicates the encryption algorithm to which the certificate owner prefers to have information encrypted by.
\end{itemize}

A PGP certificate identifies the owner of the public key and contains a signature of the key's owner, which states that the key and the identification go together. Each label contains a different means of identifying the key's owner (e.g., the owner's name and corporate e-mail account, the owner's nickname and home e-mail account, a photograph of the owner - all in one certificate). A single certificate can contain multiple signatures. Several or many people may sign the key/identification pair to attest to their own assurance that the public key definitely belongs to the specified owner. The list of signatures of each label may differ. Signatures attest to the authenticity that one of the labels belongs to the public key, not that all the labels on the key are authentic \cite{pki}. Unlike, X.509 certificate, PGP cerficates do not rely on a CA for identity-verification - but on a Web of Trust.  As this `web' grows the reputation (or binding of the certificate to the user identity therein) proportionally grows.  Since PGP allows anyone to be a `trusted introducer' on the web of trust, the identity-verification could be very subjective and easily exploited (e.g., by someone who simply signs each certificate he/she receives without verifying the requestor's identity).  This is very similar to a facebook user who accepts all friend requests without confirming whether or not they know the individual who sent the request.  Additionally, as mentioned above, because PGP has not been widely adopted and is difficult to deploy, we seek to leverage Bitcoin to encourage a larger scale adoption of PGP.  In the following subsection, we present our Bitcoin-Based PGP certificate format that builds on this concept of a web of trust to primarily overcome some of the weaknesses inherent to the PGP.

\subsection{Bitcoin-Based PGP Certificates}
Bitcoin is an experimental, decentralized digital currency that enables instant payments to anyone, anywhere in the world. Bitcoin uses peer-to-peer technology to operate with no central authority: managing transactions and issuing money are carried out collectively by the network. Bitcoin is designed around the idea of using cryptography to control the creation and transfer of money, rather than relying on central authorities \cite{bitcoinmain,bitcoinworks}.  Our Bitcoin-Based PGP certificate contains all the relevant elements found in a traditional PGP Certificate but also includes a Bitcoin Address for Identity-Verification and one used for Certificate Revocation.  A Bitcoin address is an identifier of 27-34 alphanumeric characters, beginning with the number 1 or 3, that represents a possible destination for a Bitcoin payment.  A Bitcoin transaction is a signed section of data that is broadcast to the network and collected into blocks. It typically references previous transaction(s) and dedicates a certain number of bitcoins from it to one or more new public key(s) (Bitcoin address) \cite{bitcointrans}.  Because transactions must be verified by miners, Bitcoin users are sometimes forced to wait until they have finished mining. The bitcoin protocol is set so that each block takes roughly 10 minutes to mine. In the case of a purchase, some merchants may make users wait until this block has been confirmed, which will delay the receipt of the digital goods that have been paid for - whereas in other cases (e.g., low value transactions), a merchant will give access to the goods prior to the transaction being verified by the miners \cite{coindesk}. In our case, the delay does not pose a major problem since it will only take place when a trust relationship is being established for the first time - not upon certificate generation.  The value of using Bitcoin in the context of a PGP certificate centers around the fact that because it is built upon a peer-to-peer network, it is able perform its functions (e.g., money transfers, double-spending prevention) without the aid of a CA - similar to the traditional web of trust.  This is advantageous in any context where end-to-end data confidentiality is needed or desired (e.g., email, text message, cloud sharing, or social network communications).  Users are more likely to trust an infrastructure that is independent of any CAs, but can still offer the same cryptographic guarantees (i.e., confidentiality and integrity) as an environment that is under their full control or purview. 

\section{PGP Threats and Security Goals}
In this section, we expound on the threats we identified in the introduction and describe our security goals.  We make the following assumptions as it pertains to these threats.   
\begin{enumerate}
\item The PGP users are leveraging the Web of Trust for certificate vetting.
\item The PGP user certificates are also capable of supporting a hierarchical structure and use of a CA (similar to the traditional X.509 certificate).  However, we make the assumption that a CA is not being used.   
\item PGP users can assign a level of trust to another PGP user's public key
\item PGP users can assign a level of validity to another PGP user's public key
\end{enumerate} 

\subsection {Threat Scenarios}
Although there are a number of well documented issues with PGP, we primarily focus on threats relating to certificate validation, endorsement, and trust relationship establishment.  Other threats associated with PGP, such as privacy of the Web of Trust and the fair exchange of Certificate Endorsers, are outside of the scope of our research.  One of the main issues with the Web of Trust model is that beyond a first degree trust relationship, it is difficult to quantify trust.  As a result, it becomes a scenario that is relatively easy for an attacker to exploit.  This is primarily due to the fact that within PGP, anyone can serve as a trusted introducer (essentially a CA) to another individual. This is analogous to a member of a social network accepting every friend request that he/she receives.  There is no way of knowing whether or not the individual being vouched for is actually who they say they are - which makes this a very dangerous weakness of the current PGP Web of Trust model. 

One can mitigate the threat of compromise through use of our new Bitcoin-Based PGP certificate and Public Key validation process.  Recall that during the identity-verification process, there is always the risk of losing the Bitcoins that are sent to a Certificate Owner.  However, in the event that the Bitcoins are not returned (e.g., due to an adversarial threat or greedy Certificate Owner), the more valuable commodity (the information) will be spared from being exposed.  In our new certificate, since we associate a Bitcoin Value with a trust relationship, it is easy to infer (by examining previous identity-verification transactions), how much an individual is trusted. By using Bitcoin as a form of trust establishment, as the number of transactions increase to a particular Bitcoin-Based PGP certificate, the level of trust will also increase in the associated identity.\\
\indent As mentioned above, the levels of trust with the PGP are difficult to rely on with any certainty. There are 3 levels of trust that can be assigned to someone's Public Key within PGP: Complete Trust(2), Marginal Trust(1), or No Trust (0).  Similarly, there are also 3 levels of validity: Valid, Marginal Validity, or Invalid.  PGP requires at least 1 Completely Trusted signature or 2 Marginally Trusted signatures to establish a Public Key as Valid \cite{pgpworks}.  Even with its numerical value system that maps to a particular ``level'', PGP users are subject to the loose definitions of validity and trustworthiness of a PGP key.  The highest level of trust in a key, implicit trust, is trust in a Certificate Owner's own key pair. PGP assumes that if you own the private key, you must trust the actions of its related public key \cite{pgpworks}.  However, beyond one's own key pair, how can an actual value be assigned to a trust relationship?  This challenge arises when users are asked to make security decisions. In a recent article, Michigan State University professor Rick Walsh studies the reasoning process behind the decisions people make that lead to computer security breaches \cite{poor}.  His research shows that how people visualize and conceptualize hackers and other cybercriminals affects their cybersecurity decision-making.  In the case of this threat, most users will assume that the mere fact that they are using PGP makes them secure (i.e., their visualization of the adversary).  Hence, their concern of assigning an appropriately level of validity or trust to another's PGP key is unlikely.  It is also likely the case that there is a lack of understanding of what an appropriate designation would be - unless training were provided or there was a physical trust relationship between verifier and Certificate Owner.\\     
\indent With our new endorsement process offered via Bitcoin, this threat would be mitigated by the following constructs of our scheme:  1) Certificate Signing MUST precede the incentive fee. A fixed fee of 0.001 BTC is sent to the Bitcoin address provided by the certificate endorser (fee is paid from the certificate owner's bitcoin address - as available).  This fee cannot be modified by the certificate owner OR the endorser - however, it serves as a small incentive (e.g., Thank You for Signing) to willing and competent endorsers, 2) Endorsement process is not automated. Our prototype forces users to go through a step by step process in order to sign a certificate stored on our server, and 3) Levels of Trust are established by the certificate endorser, not certificate owner. In our scheme, when performing an identity-verification transaction, any amount of Bitcoins can be sent for verification purposes.  These Bitcoins are `at risk' until the certificate owner returns them.  As a result, this serves as a very clear indication of trust between certificate endorser and owner.  For example, one would risk more Bitcoins in verifying a close friend's PGP certificate than someone who was not in his/her close circle of friends.  As a last option, subsequent endorsers of a particular PGP key can glean levels of trust very easily prior to `risking' their own Bitcoins using an identify-verification transaction - as previously discussed.\\ 
\indent A few additional threats to consider with leveraging Bitcoin as an alternative method of certificate verification are those related to rogue certificate owners, wealthy endorsers, and untrustworthy endorsers.  In the first case, a certificate owner can generate a PGP key and use it for collecting payments and never return incoming identity-verification transactions to endorsers.  To further complicate this scenario, a wealthy endorser risks very little by endorsing such users.  To address these threats, we still rely on the PGP trust model that allows for out-of-band methods of certificate verification and a web of trust.  The inference is that users will not initiate an identity-verification transaction with someone they do not already know and trust from prior interactions.  Additionally, in the case of the wealthy endorser, only one verification transaction is considered valid for a particular certificate.  Thus, their credibility (over time) will come into question as they continue to endorse untrustworthy certificates.  Lastly, we consider the scenario where endorsers are suspected of being malicious by endorsing `just for the sake of endorsing'.  Since our endorsement scheme does not invalidate - but augments - the endorsement process provided by PGP, over time a malicious endorser would be classified as someone who cannot be trusted - especially if they are endorsing both questionable and legitimate certificates.  A legitimate case to consider is someone who is a professional certificate endorser.  Someone whose professional responsibility is to endorse new certificates has their job (and reputation) to consider if they are found to be endorsing certificates that are not legitimate - over time.     

\section{Prototype Design}
In this section, we describe the implementation details of our prototype.  The prototype will be a hosted web application that will resemble a traditional Public Key Certificate server.  A Certificate or key server receives and serves existing cryptographic keys to users or other computer programs.  The keys distributed by the key server are almost always provided as part of a cryptographically protected identity certificate containing not only the key but also 'entity' information about the owner of the key. In the case of our Bitcoin-Based PGP Certificate server, the certificates will be in the OpenPGP public key format - with the additional Bitcoin addresses for identity-verification and certificate revocation as specified in section 3.  The primary motivations for creating a new certificate server are to 1) Accommodate our new Bitcoin-Based PGP certificates, 2) Enable Identity-Verification and Revocation transactions, and 3) Enable Certificate Signing Endorsements.  To faciliate these ``features'', our certificate server provides the following functions: Generate, Revoke, Verify, and Sign.  We discuss each in the following sub-sections after which we detail the design of our novel PGP Key server based on the blockchain as an area of future work.   

\subsection{Generate and Revoke}
Our Bitcoin-Based PGP certificates are generated using the openpgp javacript package - which is a javascript implementation of the OpenPGP protocol.  This standard is defined by \textbf{RFC4880}. This package provides functions to encrypt and sign your data and communication, features a versatile key management system as well as access modules for all kinds of public key directories. As it is designed for web-based devices, it can be used as an alternative to GnuPG, also known as GPG, which is a command line tool with features for easy integration with other applications \cite{gnupg}.  Each Bitcoin-Based PGP certificate will contain a set of required parameters prior to generation and items that will be automatically generated by the prototype application.  One thing to note is that our implementation does not require any modification of the original PGP certificate format.  We leverage the key server to store both the identity-verification and revocation addresses listed in the PGP comment field. 

In PGP, users can revoke their certificate if they feel like the certificate has been compromised. PGP also allows a user to designate a certificate revoker.  With PGP certificates, the user usually posts the revoked certificate on a certificate server.  To enable an easier revocation process for our Bitcoin-Based PGP certificate, we are able to take advantage of the the Bitcoin transaction block chain for revocation purposes.  The block chain is a transaction database shared by all nodes participating in a system based on the Bitcoin protocol. A full copy of a currency's block chain contains every transaction ever executed in the currency. With this information, one can find out how much value belonged to each address at any point in Bitcoin history \cite{blockchain}.  The immutable nature of the Bitcoin Transaction Block chain makes it attractive from a certificate revocation standpoint.  To revoke a certificate, we perform a Bitcoin transaction between the two Bitcoin addresses colocated in the Bitcoin-Based PGP certificate.  This transaction can only be performed by the certificate owner once authenticated. 

Key revocation is arguably the most important component of any certificate-based identification system. Our implementation deliberately forces the user to make a valid Bitcoin transaction to a legitimate Bitcoin address in his possession. Alternatively, the revocation status could be stored in \textit{OP-RETURN} fields if our decentralized approach is adopted (as explained later in the paper). 
Our current method, however, has an important technical advantage: It makes verification of a certificate status extremely efficient since revocation transactions will be stored in the Bitcoin Unspent Transaction Outputs (UXTO) database and propagated among all nodes automatically. The UXTO are redeemable transactions and the information on certificate status will be kept in main memory for efficient verification.   

\subsection{Verify and Sign}An identity-verification transaction is the primary mechanism by which a user can verify another user's Public Key in a Bitcoin-Based PGP certificate.  We accomplish this by associating a Bitcoin Address with each generated PGP certificate.  To perform the Bitcoin Identity-Verification transactions, we leverage the `jsonRPCClient' library's \textbf{sendtoaddress} function.  This function sends a specified amount from a server's available balance to a specified bitcoin address.  It takes the following parameters:\\ 

\begin {compactitem}
\item \textit{bitcoinaddress} - Bitcoin address to send to.
\item \textit{amount} - Amount to send (float, rounded to the nearest 0.01).
\item \textit{minconf} - Minimum number of confirmations req'd for transferred balance.
\item \textit{comment} - Comment for transaction. 
\item \textit{comment-to} - Comment for to-address. \\
\end {compactitem}

The first step in an identity-verifcation transaction is to connect to a local or remote bitcoin instance.  In the case of our web server, users are required to have an existing Blockchain account (or they can sign up for one via our site).   Next, the \textbf{getBalance()} function is used to  determine if there are enough funds in which to initiate an identity-verification transaction.  If so, the first transaction, `Tx-1', which represents the initial identity-verification transaction, is sent to the Certificate owner's Bitcoin Address.  Once `Tx-1' is received by the Certificate owner, the Certificate owner has the option to send the funds back to the verifier in transaction `Tx-2'.  Lastly, once `Tx-2' is received by the verifier, the transactions `Tx-1' and `Tx-2' are compared for equality.  If they are equal, the verifier can proceed to confirm the validity of the certificate and use it for regular PGP certificate operations.  Once the identity-verification transaction is complete, the verifier will have the option of signing the Public Key of the Certificate that was just verified.  This function mirrors the traditional signing of a Public Key, but differs in the fact that the signature will be added to the list of Web of Trust signatures associated with the Public Key signed.  It also results in an endorsement for the signer - which is automatically sent to the signer's bitcoin address. The entire certificate endorsement process between a user Alice and a user Bob is shown in below.  As shown, this entire process is conducted via our Bitcoin-Based PGP Key Server - making it very easy to endorse and verify new PGP Keys.  The server also supports the use of PGP keys for normal operations such as encryption and decryption of standard ASCII text.  After each identity-verification transaction and certificate signing (endorsement) operation, we provide a link to the transaction hash for external verification purposes (i.e., via the blockchain itself).

\begin{figure}[h]
\centering  \includegraphics[width=8cm,height=8cm]{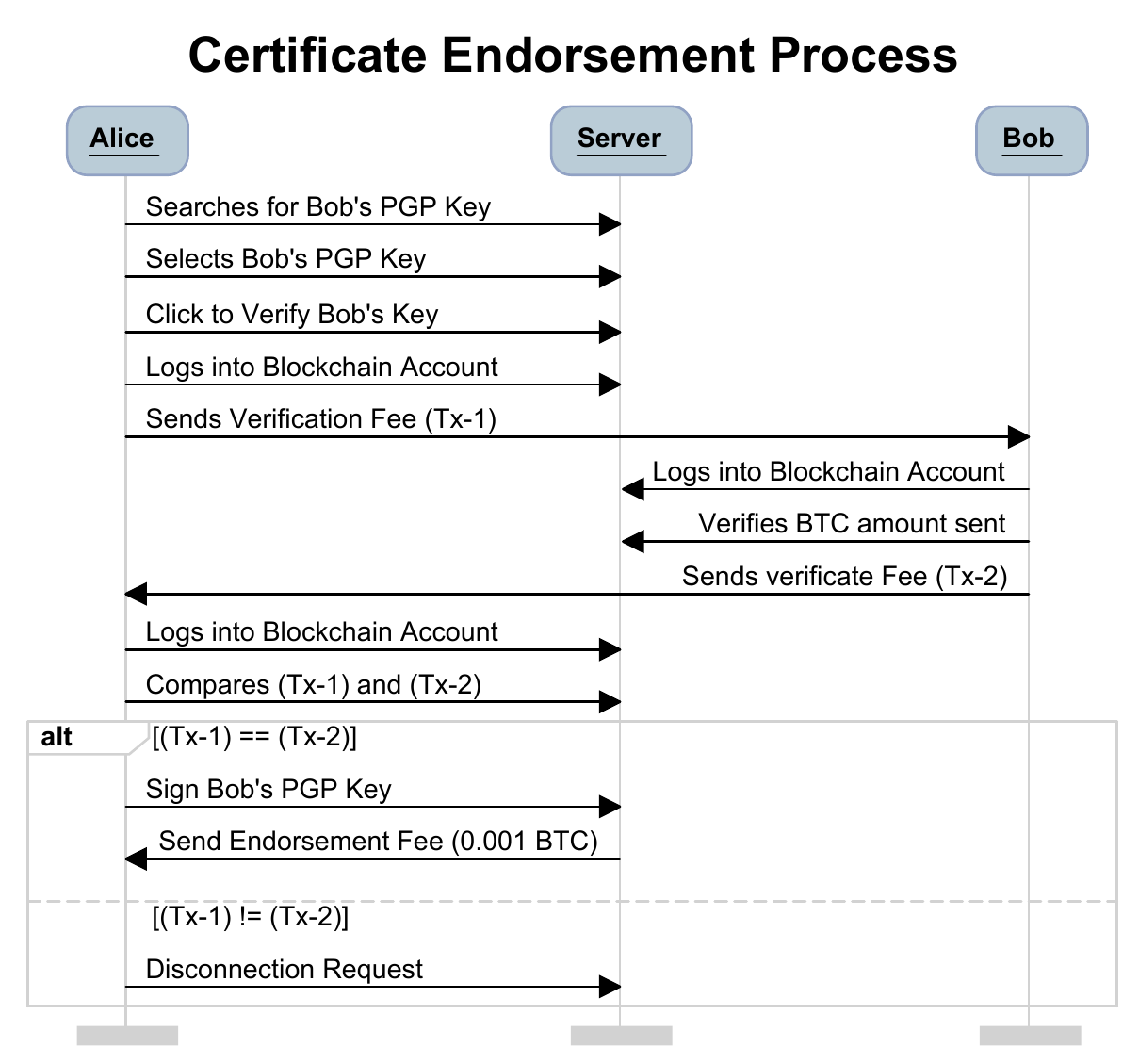}
\end{figure}

\subsection{Blockchain PGP Key Server}
In this subsection we describe the design of a novel PGP Key Server based on the Bitcoin Transaction Blockchain.  Historically, the use of bitcoin’s blockchain to store data unrelated to bitcoin payments has been a controversial subject. Many developers consider such use abusive and want to discourage it. Others view it as a demonstration of the powerful capabilities of blockchain technology and want to encourage such experimentation. Those who object to the inclusion of non-payment data argue that it causes ``blockchain bloat'', burdening those running full bitcoin nodes with carrying the cost of disk storage for data that the blockchain was not intended to carry. Moreover, such transactions create UTXO that cannot be spent, using the destination bitcoin address as a free-form 20-byte field. Because the address is used for data, it does not correspond to a private key and the resulting UTXO can never be spent \cite{chimera}. As a result, these transactions continue to increase the size of the blockchain over time.  

In version 0.9 of the Bitcoin Core client, a compromise was reached with the introduction of the \textit{OP-RETURN} operator. \textit{OP-RETURN} allows developers to add 40 bytes (now 80 bytes) of nonpayment data to a transaction output. However, unlike the use of "fake" UTXO, the 
\textit{OP-RETURN} operator creates an unspendable output, which does not need to be stored in the UTXO set. \textit{OP-RETURN} outputs are recorded on the blockchain, so they consume disk space and contribute to the increase in the blockchain’s size, but they are not stored in the UTXO set and therefore do not bloat the UTXO memory pool and burden full nodes with the cost of more expensive RAM. \cite{chimera}.  As a result, innovative decentralized applications such as the one subsequently described can be realized.  We focus our design discussions on storage and retrieval methods we would employ in an actualization of a prototype application and summarize with an overall component diagram. 

\textbf{STORAGE: } Depending on the size of PGP key generated, the size could range from ~1018 bytes (1024-Bit key) to ~3100 bytes (4096-Bit key).  PGP supports up to an 8192-Bit key which corresponds to an even larger on-disk or memory capacity for storage purposes. Keeping this in mind, along with the fact that the blockchain only accepts `data' transactions of up to 80 bytes in size, our storage leverages an innovative certificate fragmentation mechanism to enable both logical storage and efficient retrieval.  A message within our PGP Key Server will consist of a 5 Byte Header which will include the PGP Key ID (KeyID), Fragment ID (FID), Total Fragments (Total), and the Message Data.  Message format is shown below: 

\begin{figure}[h]
  \centering \includegraphics[width=8cm,height=4cm]{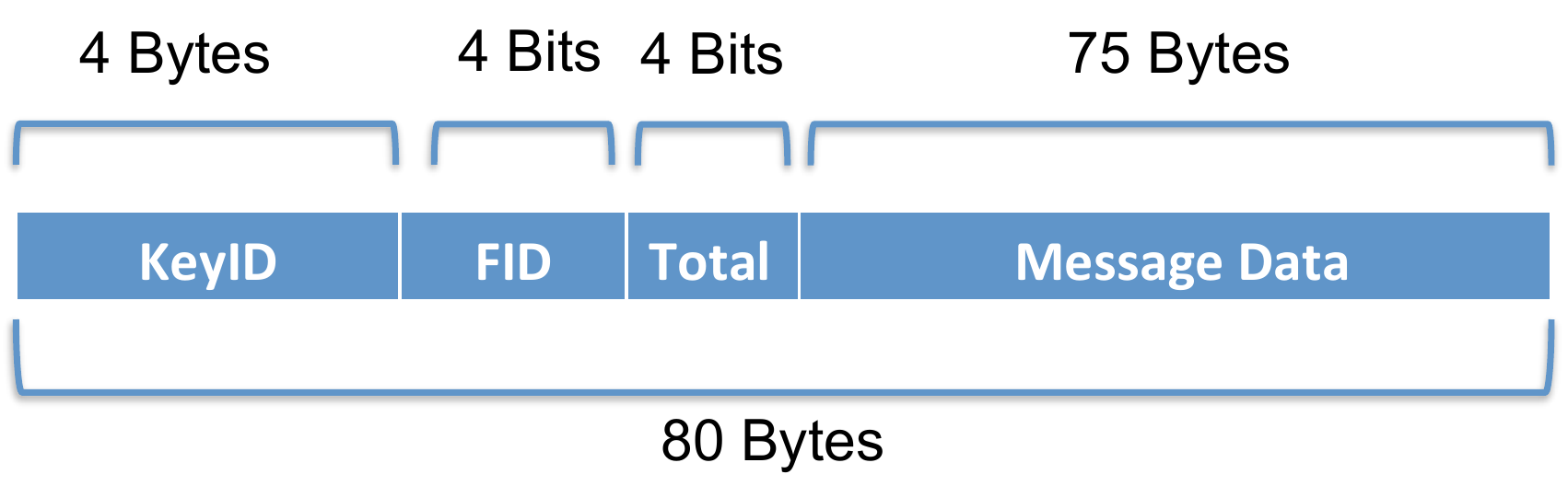}
  \caption{Blockchain PGP Message}
\end{figure}

Upon initiating a storage request, a user will provide the KeyID and PGP Key (Message Data) to the client application.  The Bitcoin Address associated with the PGP Key will also be extracted and provided as input - as it will be required during the information retrieval process.  The processing of each PGP Key will take place as follows: 
\begin{algorithm}
\KwIn{PGP-Key, PGP-Key-Length, PGP-Key-ID, Bitcoin Address}
 \KwResult{TRUE or FALSE}
 ChunkSize=75; Pointer=0\;
 Total Fragments=PGP-Key-Length / 75\;
 Index=0\;
 Result=FALSE\;
 Fee=0.001\;
 \While{(Pointer LESS THAN PGP-Key-Length)}{
  DATA = readKey (PGP-Key, Pointer, ChunkSize)\;
  \eIf{\textbf{more} DATA}{
   Header = [PGP-Key-ID + Index + Total Fragments]\;
   Send-Message (Header, DATA, Bitcoin Address, Fee)\;
   Index = Index + 1\;
   Pointer = Pointer + Chunksize\;
   Process Next Chunk (Continue)\;
   }{
   Done Processing\;
   Result=TRUE\;
  }
  Return Result;
 }
 \caption{Process PGP Key}
\end{algorithm}

In short, this algorithm computes Bitcoin transactions for the PGP Key 75 bytes at a time since this is the allowed size of our PGP Key Server messages (given the header of 5 bytes).  A few things to note: Total Fragments denote the number of fragments there will be given the size of the PGP Certificate.  In the simplest example, a PGP key of size 75 bytes will result in a Total Fragment computation of 1.  The round function (or similar) will have to be used to ensure that the Total Fragment count is a whole number.  The Pointer variable is used to both split and process each PGP Key segment from the beginning of the PGP Key to its end.  Prior to discussing the retrieval method we employ, there are a number of similar libraries in existence that are used for similar purposes (i.e., to store data in the Bitcoin transaction blockchain.  It is helpful to describe a few of the key ones here: 1) Coinspark allows you to Add messages to bitcoin transactions and Transfer any asset over the Internet, 2) Coinsecrets enables \textit{OP-RETURN} transactions to be retrieved from the bitcoin or testnet blockchains, as well as unconfirmed transactions from the memory pool. Their API also parses the content of the \textit{OP-RETURNs} for display in decoded form, and 3) Factom is leveraged by Businesses and governments to simplify records management, record business processes, and address security and compliance issues \cite {coinspark,coinsecrets,factom}.

\textbf{RETRIEVAL: } The Retrieval of a PGP Key from the blockchain requires several steps that mirrors (in reverse) the storage operations performed.  Similar to the defragmentation process of an IP datagram.  The steps are as follows: 

\begin{enumerate} 
\item User requests {Bitcoin-Address, KeyID} from Client application (Bitcoin-Address is used to identify \textit{OP-RETURN} transactions associated with KeyID.
\item Application Searches Blockchain for \textit{OP-RETURN} transactions associated with Bitcoin-Address (using the \textit{chain.com} API) and stores transactions in Results Array.
\item Application Extracts KeyID from Element[0] of Results array element
\item Application Verifies that KeyID matches KeyID from Step 1
\item Application Extracts Total Fragments from Element[0] of Results array element
\item Application Verifies number of transactions (Results Array Length) retrieved equals Total Fragments in PGP Key Message
\item If ALL transactions were retrieved successfully, Application reorders Messages and stores Messages in Ordered-Results Array
\item Reassemble PGP Key from Ordered-Results Array
\item Utilize PGP Key for Normal Operations (Sign, Verify, Encrypt, Decrypt)
\end{enumerate}

Chain's enterprise-grade block chain API makes it easy to build Bitcoin applications that are fast, reliable, and secure \cite{chain}.  The Get Bitcoin Address \textit{OP-RETURN} function returns any \textit{OP-RETURN} values sent and received by a Bitcoin Address in an Array of \textit{OP-RETURN OBJECTS}.  The \textit{OP-RETURN OBJECT}  is a pseudo-object that is extracted from the Transaction Object. It is an interpretation of the \textit{OP-RETURN} script within a zero-value output in a Bitcoin transaction. The \textit{OP-RETURN} can be used to include 80 bytes of metadata in a Bitcoin transaction.  Each \textit{OP-RETURN OBJECT} Contains the following data that we leverage for step 2 above.  Lastly, it is important to note that the \textit{OP-RETURN} data will have to be decoded prior to being used.  This decoding process is elegantly handled by the chain API.  This decoded text is the contents of our message that we specify in step 2 above.  We conclude this section with a overview diagram (Figure 2) of how each PGP message fragment looks within the overall structure of the Bitcoin Transaction Blockchain.  It is highly probable that each message could be contained in a disparate block within the transaction blockchain.  Thus the header is vital in ensuring the retrieval process works seamlessly.  In Figure 2, we demonstrate a sample message broken into six message fragments and stored in three separate transaction blocks. 

\begin{figure}[h]
  \centering
  \captionsetup{justification=centering}
  \caption{PGP Message Stored in Transaction Blockchain}
  \includegraphics[width=8cm,height=4cm]{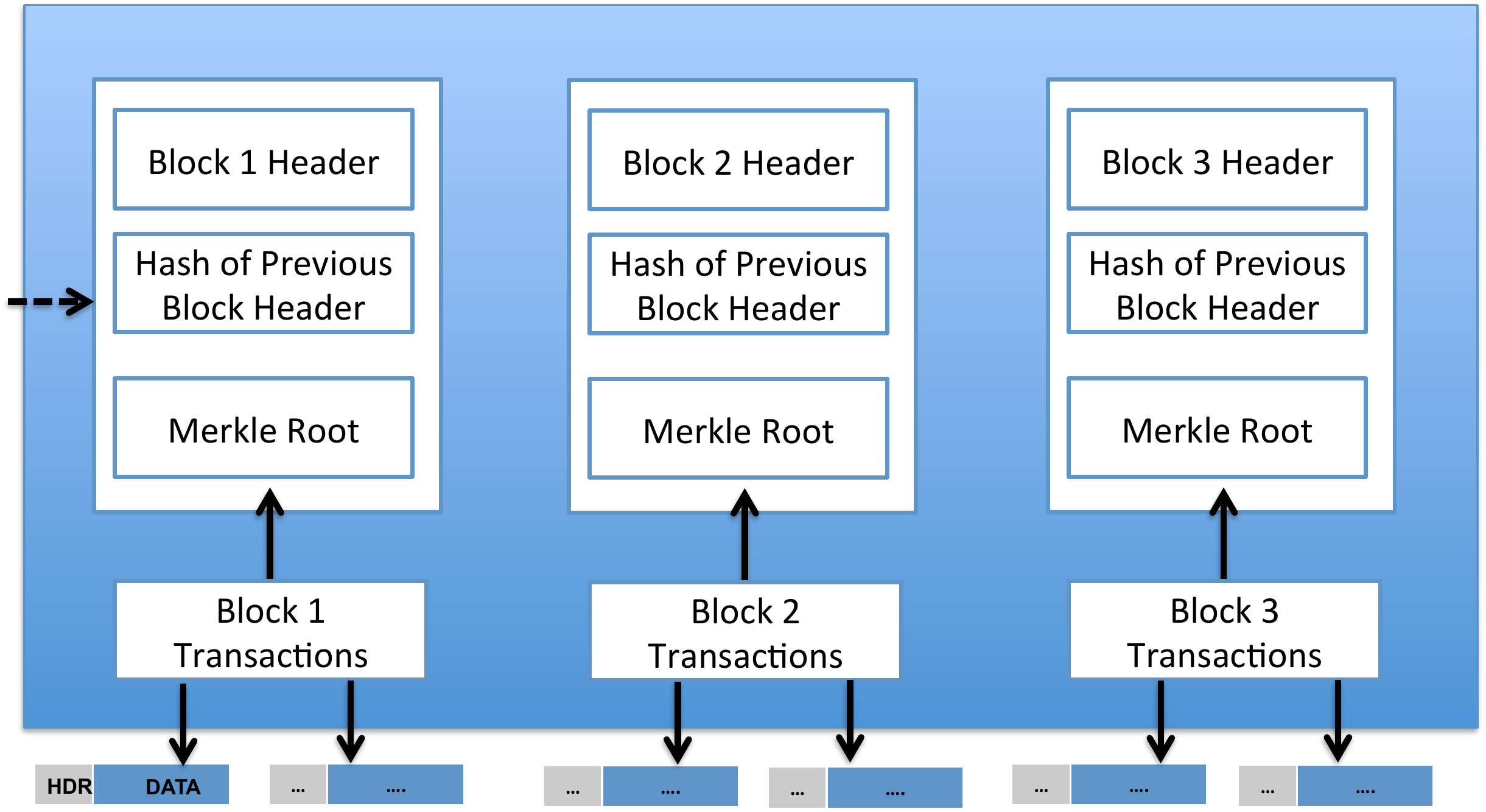}
\end{figure}

\section{Prototype Demonstration Output}
In this section, we walk through the steps of using our prototype application to perform the bitcoin identity-verifications functions and the output produced by our application at each stage (as applicable).  

\begin{enumerate}
\item Alice (Certificate Owner) Generates Bitcoin-Based PGP Certificate:
\begin{verbatim}
Key-Type: RSA
Name-Comment: 
1Lk3XuR3dvPebRS6QgmVXVBjm7NBkuTuM7|
19X3kcrYNFhaPdwpdwnnBrDSNtecxUqDrW
Passphrase: pwd
Name-Real: Alice
Name-Email: alice@bitcoinpgp.com
Key-Length: 2048
%commit
\end{verbatim}

\item Alice sends Certificate to Bob (no output - performed out of band)
\item Bob sends Identity-Verification transaction to embedded Bitcoin Address
\begin{verbatim}
Current Balance is: 0.033673
How much would you like to send for 
VERIFICATION TRANSACTION? 0.00856179
Sending 0.00856179 to 
1Lk3XuR3dvPebRS6QgmVXVBjm7NBkuTuM7
\end{verbatim}

\item Alice sends Identity-Verification funds back to Bob's Bitcoin address
\begin{verbatim}
Current Balance is: 0.01198579
How much would you like to send for 
RETURN VERIFICATION TRANSACTION? 
0.00856179 Sending 0.00856179 to
1HiLRA7pC3d6mLGUpuhx3qH4G8sDVbdpD2
\end{verbatim}

\item Bob checks for return of Identity-Verification funds

\begin{verbatim}
Amount Sent: 0.00856179
Transaction of 0.00856179 Found!
Result = True
\end{verbatim}

\item Bob checks Alice's certificate for validity (result true)
\item Alice (at later date) revokes her certificate
\begin{verbatim}
Sent Revocation Transaction of Amount: 
0.000017 to Revocation Address: 
19X3kcrYNFhaPdwpdwnnBrDSNtecxUqDrW
1Lk3XuR3dvPebRS6QgmVXVBjm7NBkuTuM7 
Certificate Successfully Revoked
\end{verbatim}

\item Bob checks Alice's certificate for validity
\begin{verbatim}
Bitcoin Address Verified
Amount Sent: 0.000017
Revoke Transaction of 0.000017 Found!
\end{verbatim}

\end{enumerate}

\section{Conclusions and Future Work}
In this paper we presented a number of enhancements to PGP and associated Web of Trust - which has suffered from a littany of issues since its inception.  Specific issues of certification, endorsement, and ambiguous levels of trust have prevented its wide scale adoption.  The contributions we discuss include a novel Bitcoin-Based PGP Certificate format, Design of a Distributed PGP Key Server - using the Bitcoin transaction blockchain for certificate storage and retrieval, Certificate Signing Endorsements, and Identity-verification and revocation transactions using Bitcoin.  We demonstrate how these added features and capabilities can potentially result in a greater use of PGP which has inherently proven security properties.  Future work will consist of examining alternative methods of employing Bitcoin for identity-verification using actual Bitcoin Distributed Contracts or alternative methods that do not require modification of the original PGP certificate format.  Keybase.io allows you to get a public key, safely, starting just with someone's social media username(s), but also provides other mechanisms of verifying a particular key (e.g., pgp fingerprint and bitcoin addresses) \cite{keybase}.  In other words, it provides a compendium of online identifiers for a particular owner of a public key stored on their servers.  A potential area for future work would be to enable verifiers to leverage one or more of the online identifications provided by Keybase.io to strengthen the trust of certificate stored on our server (via their API).  Additionally, the integration of Bitcoin-Based PGP Certificates into infrastructures where secure sharing is offered (via text messaging, chat applications, and Secure Cloud Storage servers) would demonstrate their usefulness in actual environments.  Lastly, a stronger form of certificate revocation should be explored that builds on the procedure we present in our research.  We are currently working on a prototype of the Distributed Key Server that will also address some of the privacy issues found in the traditional PGP Key Servers (e.g., MIT).  Our Enhanced PGP Key Server is currently hosted at [http://enhancedpgp.com/].

%
%

\end{document}